\documentclass[reprint,aps,pra,twoside]{revtex4-1}
\usepackage[ps2pdf]{hyperref}

\usepackage{amsmath,amssymb}
\usepackage{units}
\usepackage{dcolumn}
\usepackage{graphicx}
\usepackage[bf]{subfigure}
\usepackage{pgf}

\usepackage{mathtools}

\newcommand{\res}{\mathrm{r}}
\newcommand{\q}{\mathrm{q}}
\newcommand{\dr}{\mathrm{d}}

\newcommand{\nch}{{n_\mathrm{ch}}}

\newcommand{\onoff}{{1/0}}

\newcommand{\sigmaz}{\sigma_\mathrm{z}}

\newcommand{\bi}{\mathrm{I}}
\newcommand{\bii}{\mathrm{II}}

\newcommand{\tm}{{t_\mathrm m}}
\newcommand{\Tm}{{T_\mathrm m}}
\newcommand{\taum}{{\tau_\mathrm m}}
\newcommand{\thr}{\mathrm{th}}

\newcommand{\SNR}{\mathrm{SNR}}
\newcommand{\Dopt}{D^\mathrm{opt}}
\newcommand{\Dhi}{D^\mathrm{hi}}
\newcommand{\Dlo}{D^\mathrm{lo}}

\usepackage{amsmath}

\newcommand{\bra}[1]{\left\langle{#1}\right|} 
\newcommand{\ket}[1]{\left|{#1}\right\rangle} 

\newcommand{\const}{\text{const}}

\newcommand{\slashfrac}[2]{\raisebox{.0em}{$#1$}\big/\raisebox{-.2em}{$#2$}}
\newcommand{\at}[2]{\left.{#1}\right|_{#2}}

\DeclareMathOperator\erf{erf} 

\begin{document}

\title{Optimal conditions for high-fidelity dispersive readout of a qubit \mbox{with a photon-number-resolving detector}}
\author{Andrii~Sokolov}
\email[E-mail: ]{andriy145@gmail.com}
\affiliation{Institute of Physics of the National Academy of Sciences, pr. Nauky 46, Kyiv-28, MSP 03028, Ukraine}

\begin{abstract}
We determine the optimal parameters for a simple and efficient scheme of dispersive readout of a qubit.
Depending on the qubit state (ground or excited), the resonance of a cavity is shifted either to the red or to the blue side.
Qubit state is inferred by detecting the photon number transmitted through the cavity.
It turns out that this kind of detection provides better measurement fidelity than the detection of the presence or absence of photons only.
We show that radiating the cavity on either of the frequencies it shifts to results in a suboptimal measurement. 
The optimal frequency of the probe photons is determined, as well as the optimal ratio of the shift to the resonator leakage.
It is shown that to maximize the fidelity of a long-lasting measurement, it is sufficient to use the parameters optimizing the signal-to-noise ratio in the photon count.
One can reach 99\% fidelity for a single-shot measurement in various physical realizations of the scheme.
\end{abstract}

\maketitle

\section{Introduction}

Dispersive qubit readout is a promising scheme that can be used for quantum computation.
The readout can be made fast and accurate~\cite{jeffrey2014fast}, which enables its usage in quantum error-correction schemes~\cite{nielsen2010quantum}.
It is also highly quantum-non-demolition~\cite{braginsky1996quantum} method.
Hence it can be used for a continuous-feedback correction~\cite{ahn2002continuous}, which is an appealing scheme due to a small number of requirements.
Also, it is possible to use the dispersive readout for the qubit initialization by measurement; this was done in Refs.~\cite{riste2012initialization,johnson2012heralded}.
Lastly, the properties of the dispersive measurement are advantageous to read out results of a computation.
It was recently used for readout in a digital quantum simulation of spin systems~\cite{salathe2015digital}.

The idea of the readout lies in the following.
During the measurement, a qubit is coupled to a cavity off-resonantly.
This results in the absence of energy exchange between them~\cite{sete2014purcell,berman2012dynamics}.
However, the resonant frequency of the cavity gets shifted either to the blue or to the red side, depending on  whether the qubit is in the excited or in the ground state.
The shift can be used to infer the qubit state.
To do this, one can monitor the phase quadrature of the transmitted or the reflected radiation~\cite{blais2004cavity,schuster2007circuit}.
Alternatively, one measures the intensity of the transmitted signal.

The latter possibility offers distinct advantages while being feasible for readout in both optical and microwave ranges.
In optics, it is natural to use a photon counter for detection.
It can be used for dispersive measurement in optical realizations of cavity QED~\cite{hennessy2007quantum,brennecke2007cavity}.
It is especially interesting to use the scheme for continuous monitoring of Bose-Einstein condensate in a cavity~\cite{brennecke2007cavity}.
As for the microwave domain, the readout is usually carried out with a homodyne~\cite{blais2004cavity,schuster2007circuit}.
However, advances were made in the microwave photodetection, both cryogenic~\cite{chen2011microwave,fan2014nonabsorbing,koshino2015theory} and room-temperature~\cite{divochiy2008superconducting}.
It possible to use the detectors for the dispersive readout in the microwave cavity QED.
Due to its simplicity, the readout with a photon counter is appealing for scaling the measurement to multiple qubits and for integration on chip.
It is especially suited for the integrated superconducting analogues of cavity QED~\cite{blais2004cavity,schuster2007circuit}.
There, it is unclear how to realize an on-chip homodyne, as it is challenging to isolate a strong local-oscillator pump from the rest of the chip circuitry.
The photodetection scheme is free of such issues.

The use of a photodetector for the dispersive readout was already investigated in Ref.~\cite{govia2014high}.
There, the detector absorbs the cavity photons when the cavity is already populated.
We consider the case of continuous drive and permanent resonator leakage, which is easier to achieve experimentally.
Moreover, in the reference, the click/no-click detector is studied: that is, a detector that can't provide any information besides presence or absence of photons.
In contrast, we deal with a detector able to distinguish any number of incident photons.
We will show that, in the setup we consider, such a detector allows one to reach higher measurement accuracy.

In this paper, we study the performance of a simple scheme for dispersive readout with a photodetector.
Readout accuracy can be characterized with a probability of correct measurement result, i.e. fidelity.
We find the measurement parameters maximizing fidelity: the drive-resonator detuning and the ratio of a pull in the cavity resonance $g\lambda$ to the resonator leakage rate.
The drive frequency is usually taken to match the pulled cavity resonance~\cite{berman2011influence,berman2012dynamics,govia2014high}.
However, we show this may result in a suboptimal fidelity.
We find the optimal detuning and resonator leakage.
Surprisingly, they vary with the measurement duration.
The other approach is to estimate the readout accuracy with a signal-to-noise ratio (SNR) in the photon count.
Maximization of SNR yields the parameters which are constant and simpler to use.
The circumstances when these parameters result in close-to-optimal fidelity are determined.
We use our findings to estimate the duration of a high-fidelity, single-shot measurement for different physical realizations of the scheme.

\section{Measurement scheme}
\label{seqScheme}
\begin{figure}
\centering
\includegraphics{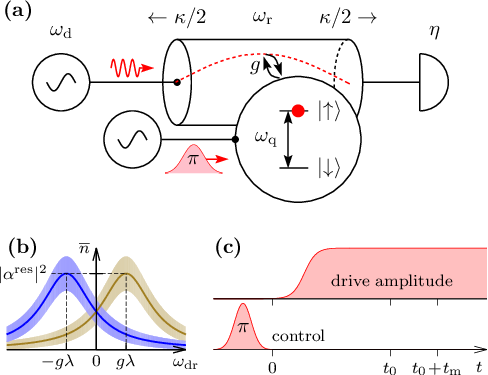}%
\subfigure{%
	\label{figSystem}%
}
\subfigure{%
	\label{figTransmission}%
}%
\subfigure{%
	\label{figSeq}%
}%
\caption{(Color online) \subref{figSystem} Schematic of the system which consists of: generators, a resonator coupled to a qubit, and a photocounter.
	\subref{figTransmission} Average number of photons transmitted as a function of the drive-cavity detuning, for the qubit in the ground state (blue line) and in the excited state (brown line).
	The root-mean-square deviation in the photon number is shown by the fill of respective color.
	\subref{figSeq} Measurement sequence (times not to scale).
}
\end{figure}

We consider the following system~[see Fig.~\ref{figSystem}].
A qubit interacts with one of resonator modes.
From the one side, the resonator is driven with a classical quasi-monochromatic pump.
On the other side, a photon-number-resolving detector is placed.
Photons leak out from the resonator by both sides.
Both the detector and the drive source do not reflect the photons.
Either of them may be connected to the cavity by means of waveguides.
Alternatively, the detector may be coupled directly to the resonator~%
\footnote{As the detector is fully-absorbing, it can be modeled by an infinite waveguide matching the impedance of a real one.
Both cases of direct and by-waveguide coupling are then modeled in the same way.}.
The qubit state is controlled by a separate line.

The system is operated in the dispersive regime.
The regime is set by the condition
\begin{gather}
\label{eqDispersive}
	(\nch+1)/n_\text{cr} \ll 1, \quad n_\text{cr} = (2\lambda)^{-2},
\\
\label{eqLambda}
	\lambda = g / (\omega_\q - \omega_\res).
\end{gather}
Here $\nch$ denotes the characteristic number of photons occupying the resonator.
$\nch+1$ is the maximal number of excitations in the qubit-resonator system, which is reached when the qubit is excited.
$n_\text{cr}$ is termed the critical photon number:
the dispersive regime breaks down at this number of excitations.
$\omega_\q$ is the qubit transition frequency, $\omega_\res$ is the resonant frequency of the cavity, $g$ is the qubit-resonator coupling.
The frequencies are bare, that is, determined for the qubit-resonator interaction turned off.
Under the condition~\eqref{eqDispersive}, it is unlikely for the qubit and the cavity to exchange an excitation: there are relatively few of them while the interaction is weak.
The effect of the coupling is only to shift the qubit and the resonator frequencies.
Here we are interested in the cavity resonance $\tilde\omega_\res$.
For the qubit in one of its eigenstates,
\begin{equation}
	\tilde\omega_\res = \omega_\res + g\lambda\sigmaz,
\quad
	\sigmaz = \pm 1.
\end{equation}
As shown in Fig.~\ref{figTransmission}, the cavity transmission then depends on a qubit state.
By counting the transmitted photons, one can perform a non-demolition readout of the qubit.
More details can be found, for example, in Refs.~\cite{berman2011influence,blais2004cavity,boissonneault2009dispersive,haroche1999cavity}.

We now describe the measurement sequence~[see Fig.~\ref{figSeq}].
First the system is thermalized.
Qubit relaxes into the ground state.
In case the measurement is to be carried out with the excited qubit, a $\pi$-pulse~\cite{martinis2003decoherence} is applied to the qubit at time $t = 0$.
At the same time, the resonator drive is turned on.
Some time after that, at $t = t_0$, one begins to count the detected photons.
The time is chosen so that all transients in the cavity have faded and the drive can be considered monochromatic.
The sequence ends at $t = t_0 + \tm$.
It can be repeated to decrease the probability of an erroneous inferring of the qubit state.

\section{Photocounting statistics}
To characterize the measurement, we need the dependence of photocounting statistics on the qubit state.
We restrict ourselves to the simplest case of the qubit occupying one of its eigenstates.

The system is well-studied in the literature.
Below, we outline the results necessary for obtaining the statistics.

The state of the resonator field is known.
The full system state is found in Sec.~V~B of Ref.~\cite{gambetta2006qubit}.
It follows that in the steady-state case, the resonator field is in a coherent state $\ket{\alpha_{\uparrow,\downarrow}}$ with either of the amplitudes
\begin{gather}
\label{eqAlphaExcited}
	\alpha_\uparrow = \epsilon / (\omega_{\dr\res} - g\lambda + i\kappa/2),
\\
\label{eqAlphaGround}
	\alpha_\downarrow = \epsilon / (\omega_{\dr\res} + g\lambda + i\kappa/2)
\end{gather}
depending on whether the qubit resides in the excited $\ket\uparrow$ or the ground $\ket\downarrow$ state.
Here $\epsilon$ is a quantity proportional to the drive amplitude, $\omega_{\dr\res}$ denotes the drive-resonator detuning, $\kappa$ is the damping rate of the resonator.
Eqs.~(\ref{eqAlphaExcited}-\ref{eqAlphaGround}) can be understood in a simple way.
The qubit and the resonator are coupled dispersively, which means they don't exchange energy.
Apart from the shift in its resonance, the cavity can be considered as not interacting with the qubit.
A standalone resonator driven by a continuous monochromatic pump is known to reside in the coherent state~\cite{mandel1995optical}.
The amplitude of that state is determined by the drive-resonator detuning, which depends on the qubit.

A correspondence is known between the resonator field and the field incident on the detector, which makes it possible to determine the photocounting statistics.
According to the input-output relations~\cite{walls2007quantum}, the output field at the second port depends linearly on the resonator field: $b^\bii_\text{out} = i\sqrt{\kappa/2} \, a + b^\bii_\text{in}$.
Note we have considered the cavity to be symmetrical.
We assume a radiation in the vacuum state $\ket{0^\bii_\text{in}}$ incident on the second port.
The probability to detect $n$ photons for the qubit in either of the eigenstates $q = \, \uparrow, \downarrow$ is given by the Mandel's formula,
\begin{equation}
\label{eqMandel}
	P_q = \bra{\alpha_q, 0^\bii_\text{in}} 
			: W^n \, e^{-W} / n! :
		\ket{\alpha_q, 0^\bii_\text{in}}
\end{equation}
where $W = \eta \int_{t_0}^{t_0 + t_m} dt b^{\dag\bii}_\text{out} b^\bii_\text{out}$, while $\eta$ is the detector efficiency, $t_0$ is the moment one begins to count photons, and $t_m$ is the time of counting.
This reduces to
\begin{equation}
\label{eqP}
	P_q(n) = \frac{(\overline n_q)^n}{n!} e^{-\overline n_q}
\end{equation}
with
\begin{equation}
\label{eqMeanCount}
	\overline n_q = \eta \tm |\alpha_q|^2 \slashfrac {\kappa} 2
\end{equation}
the average count of detected photons.
The mean square deviation of $n$ is:
\begin{equation}
\label{eqMeanSquaredCount}
	\overline{\Delta n^2}_q = \overline n_q.
\end{equation}
We express the average count~\eqref{eqMeanCount} in terms of the system parameters explicitly. 
Substituting~(\ref{eqAlphaExcited}-\ref{eqAlphaGround}) into~\eqref{eqMeanCount} and rearranging, one has
\begin{equation}
\label{eqMeanCountExplicit}
	\overline n_\uparrow =
		\frac{\eta t_m (\slashfrac{\kappa}2)^3 \, |\alpha^\mathrm{res}|^2}
			{(\omega_{\dr\res} - g\lambda)^2 + \kappa^2/4},
\quad
	\overline n_\downarrow =
		\frac{\eta t_m (\slashfrac{\kappa}2)^3 \, |\alpha^\mathrm{res}|^2}
			{(\omega_{\dr\res} + g\lambda)^2 + \kappa^2/4}
\end{equation}
with
\begin{equation}
\label{eqResonantNumberOfPhotons}
	|\alpha^\mathrm{res}|^2 = \slashfrac{4\epsilon^2}{\kappa^2}
\end{equation}
the average number of photons that enter the cavity at resonance.

By~\eqref{eqMeanCountExplicit}, statistics~\eqref{eqP} has symmetries: $\omega_{\dr\res} \to -\omega_{\dr\res}$, `$\uparrow$' $\to$ `$\downarrow$' and analogous for $g\lambda$.
This is a consequence of neglecting relaxation.
Readout characteristics are set by the statistics;
as they can't depend on qubit states labeling, they are even functions of $\omega_{\dr\res}$ and $g\lambda$.
Hence it is enough to consider the case of positive detuning and pull,
\begin{equation}
\label{eqPositiveDetuningPull}
	\omega_{\dr\res}, g\lambda > 0.
\end{equation}

Note that the results presented above rely on the rotating-wave approximation.
This imposes a restriction on the qubit and resonator frequencies:
\begin{equation}
\label{eqRWA}
|\omega_\res - \omega_\q| \ll \omega_\res + \omega_\q.
\end{equation}

Let us discuss when it is possible to consider the resonator to be in the steady state.
First, all transients in the resonator should vanish before the measurement begins.
The condition for that is
\begin{equation}
\label{eqSteadyMeasurement}
	t_0 \gg \kappa^{-1}.
\end{equation}
Second, during the measurement, the qubit should remain in the state it was set up.
In the first-order approximation in $\lambda$, the measurement does not affect the qubit occupying one of its eigenstates.
However, taking account of the next orders in $\lambda$ shows the qubit excitation can leak to the waveguides through the resonator~\cite{berman2011influence,boissonneault2009dispersive,sete2014purcell}.
Moreover, our model does not account for relaxation sources other than the waveguides~(see Ref.~\cite{muller2015interacting}, for example).
In the present work it is sufficient to characterize all those processes with the longitudinal relaxation time $T_1$.
Then the condition we have been talking about reads
\begin{equation}
\label{eqNonDemolitionMeasurementSingleShot}
	t_0, \tm \ll T_1.
\end{equation}

The condition~\eqref{eqNonDemolitionMeasurementSingleShot} is given in assumption of the single-shot measurement.
However, it changes if one is free to perform a sequence of short measurements.
In this case, one collects photons in $N$ bins, each lasting $\tm / N$;
a bin is carried out with a `fresh' qubit, prepared in a given state.
The sum of all photons collected in bins obeys the same formula~\eqref{eqMandel}.
Therefore, our arguments apply for the case of sequential measurement, with $\tm$ denoting the sum of bin durations.
An analog of~\eqref{eqNonDemolitionMeasurementSingleShot} for this case is
\begin{equation}
\label{eqNonDemolitionMeasurementSequential}
	t_0, \tm / N \ll T_1.
\end{equation}

Having the photocounting statistics, in the next two sections we characterize the robustness of the qubit measurement.

\section{Signal-to-noise ratio}

In this section, we calculate SNR of a count of detected photons, and the conditions optimizing the ratio.
The conditions for the maximal SNR is determined for two cases.
First, we suppose one has a measurement setup with a fixed dispersive pull $g\lambda$ and a fixed resonator damping rate $\kappa$.
With respect to $g\lambda$ and $\kappa$, an optimal detuning $\omega_{\dr\res}$ is found.
In the other case, either $\kappa$ or $g\lambda$ can be varied as well.

We define SNR in our measurement as follows.
A useful signal is the difference between the average photocounts for the qubit in the excited and the ground states.
In both cases, the number of the photons detected fluctuates around its mean [see Fig.~\ref{figTransmission}].
Noise in the signal is then given by the sum of fluctuations in both cases.
That is,
\begin{equation}
\label{eqSNRGeneral}
	\SNR = \frac {|\overline n_\uparrow - \overline n_\downarrow|}
		{\sqrt{\overline{\Delta n^2}_\uparrow}
			+ \sqrt{\overline{\Delta n^2}_\downarrow}}.
\end{equation}
A somewhat different expression is given in Ref.~\cite{fan2014nonabsorbing} in a similar context.
While there is no substantial quantitative difference in using those two, for our purposes the form~\eqref{eqSNRGeneral} results in cleaner math.
Particularly, it is this form that appears naturally in the expression for fidelity in Sec.~\ref{secGaussian}.

For Poissonian statistics we have,~\eqref{eqSNRGeneral} simplifies.
Substituting~\eqref{eqMeanSquaredCount} into it and rationalizing denominator gives rise to
\begin{equation}
\label{eqSNRGeneralPoissonian}
\SNR = \sqrt{\overline n_\uparrow} - \sqrt{\overline n_\downarrow}.
\end{equation}
While deriving the last expression, we have used that $\overline n_\uparrow > \overline n_\downarrow$ due to~\eqref{eqPositiveDetuningPull}.
A different expression for SNR is widely used in the literature on cavity quantum electrodynamics with superconducting circuits~\cite{gambetta2006qubit,gambetta2008quantum,schuster2007circuit,boissonneault2008nonlinear}.
It is obtained for the homodyne measurement.

We proceed to determine the conditions of maximum of SNR given by this expression.

\subsection{Maximizing SNR with respect to detuning}
\label{secSNRoptDX}
Let us consider the case of fixed resonator damping rate $\kappa$ and fixed dispersive pull $g\lambda$.
Also, we require that the average number of photons that would dwell in the cavity at resonance, $|\alpha^\mathrm{res}|^2$~\eqref{eqResonantNumberOfPhotons}, is maintained constant.
With a varying detuning, this can be achieved by the appropriate choice of the drive power, which changes $\epsilon^2$ proportionally.
Under these circumstances, we find the detuning that maximizes SNR.

Here it is convenient to introduce a set of dimensionless notations.
With the notations, the mean counts~\eqref{eqMeanCountExplicit} are expressed as
\begin{equation}
\label{eqMeanCountDX}
	\overline n_\uparrow = \frac{\taum}{(D-X)^2 + 1},
\quad
	\overline n_\downarrow = \frac{\taum}{(D+X)^2 + 1},
\end{equation}
where
\begin{equation}
\label{eqTaum}
	\taum = \eta \frac \kappa 2 |\alpha^\mathrm{res}|^2 \tm
\end{equation}
is the dimensionless measurement time.
$\taum$ gives the average count of photons absorbed by the detector in case the drive is resonant with the cavity.
Also,
\begin{equation}
\label{eqDX}
	D = \frac{\omega_{\dr\res}}{\kappa/2},
\quad
	X = \frac{g\lambda}{\kappa/2}
\end{equation}
are the dimensionless detuning and dispersive pull.

With~\eqref{eqMeanCountDX}, Eq.~\eqref{eqSNRGeneralPoissonian} takes the form
\begin{equation}
\label{eqSNR}
	\SNR = \sqrt{\taum} \left(
		\slashfrac 1 {\sqrt{(D-X)^2 + 1}} - \slashfrac 1 {\sqrt{(D+X)^2 + 1}}
	\right).
\end{equation}

Now we find the optimal detuning $\Dopt$.
Carrying out the derivative of~\eqref{eqSNR} and equating it to zero one arrives at
\begin{equation}
\label{eqDOptimal}
	\frac{\Dopt+X}{\big[(\Dopt+X)^2 + 1\big]^{3/2}}
		- \frac{\Dopt-X}{\big[(\Dopt-X)^2 + 1\big]^{3/2}} = 0.
\end{equation}
Due to~\eqref{eqPositiveDetuningPull}, $\Dopt,X>0$. 
Therefore, the equation could have a real solution only if
\begin{equation}
\label{eqDOptimalExists}
	\Dopt > X.
\end{equation}
One can check that a critical point from Eq.~\eqref{eqDOptimal} is indeed a point of maximum.
Another inequality,
\begin{equation}
\label{eqDOptimalIsAMaximum}
	\Dopt < \sqrt{X^2 + 1},
\end{equation}
is of use.
It can be obtained by multiplying the left- and the right-hand sides of the inequality $\sqrt{\Dopt+X} > \sqrt{\Dopt-X}$ by the first and the second term from Eq.~\eqref{eqDOptimal} and performing some algebra.
It is difficult to obtain an analytical solution to Eq.~\eqref{eqDOptimal}.
We have found numerically its roots in the range given by~\eqref{eqDOptimalExists} and~\eqref{eqDOptimalIsAMaximum}.

\begin{figure}[ht]
\includegraphics{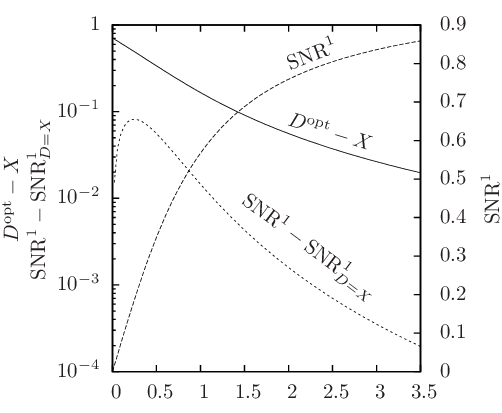}
\caption{
	Comparison of the optimal-SNR~($D=\Dopt$) and the naive~($D=X$) drive-resonator detuning.
	$\SNR^1$ denotes the SNR for $\taum = 1$.
	As defined in Eq.~\eqref{eqDX}, $D$ and $X$ are the dimensionless detuning and dispersive pull, respectively.
}
\label{figMaxSNR}
\end{figure}

In Fig.~\ref{figMaxSNR}, the difference $\Dopt - X$ is plotted and the effect of the optimal choice of detuning on SNR is illustrated.
It is seen that only at large $X$ the optimal detuning is $D \approx X$.
The difference between the optimal~($D = \Dopt$) and the naive~($D = X$) detuning is appreciable for the whole range of $X$ given in the plot;
in contrast, the interval where the increase in SNR is noticeable is substantially narrower.
It is worth using the optimal $D$ under the conditions of weak measurement, $X < 1$, as for a stronger measurement the increase in SNR is less than 1\%.

Let us obtain an approximation of $\Dopt$ for big $X$.
We expand the equation for $\Dopt$~\eqref{eqDOptimal} in series in
\begin{equation}
	\xi = \Dopt - X,
\end{equation}
leaving the terms up to the linear one.
Assuming $\xi \ll 4X$, one obtains
\begin{multline}
\label{eqFirstTermTalyorExpansion}
	\frac 1 {\big[(2X+\xi)^2 + 1\big]^{3/2}}
		= \frac 1 {(4X^2 + 1)^{3/2}}
			\left(1 - \frac 3 2 \frac {4X\xi}{4X^2 + 1}\right)
\\
		+ o\left(\frac{4X\xi}{4X^2 + 1}\right).
\end{multline}
Substituting the expression into~\eqref{eqDOptimal} and omitting the terms quadratic in $\xi$ one can solve the resulting linear equation for $\xi$.
This gives
\begin{gather}
\label{eqDMinusXApprox}
	\xi \approx \frac {2X(4X^2 + 1)} {8X^2 + (4X^2 + 1)^{5/2} - 1},
\\
\label{eqDMinusXApproxConditions}
	\xi^2 \ll [X + (4X)^{-1}]^2 / 16,
\quad
	\xi^2 \ll 1.
\end{gather}
Expression~\eqref{eqDMinusXApprox} has the anticipated asymptote,
\begin{equation}
\label{eqDOptimalAsymptote}
	\Dopt - X \underset {X\to\infty} \longrightarrow 0.
\end{equation}
From~\eqref{eqDMinusXApprox}, a simpler approximation can be obtained,
\begin{equation}
	\Dopt \approx X + 2X (4X^2 + 1)^{-3/2}.
\end{equation}
For practical purposes it works well for $X > 0.4$.

Now consider the case of small $X$.
We assume that $D > X$.
An approximate identity takes place, up to linear terms in $X$:
\begin{equation}
\label{eqFractionsSmallX}
	\frac D {\big[(D \pm X)^2 + 1\big]^{3/2}}
	\approx \frac D {(D^2 + 1)^{3/2}}
		\left(1 \mp \frac {3DX} {D^2 + 1} \right).
\end{equation}
We substitute~\eqref{eqFractionsSmallX} into~\eqref{eqDOptimal} and find the solution of the resulting equation.
This gives the small-$X$ approximation,
\begin{equation}
	\Dopt \approx \slashfrac 1 {\sqrt 2}, \quad X^2 \ll 1.
\end{equation}

SNR is determined by the interplay of two effects that depend on the resonator decay rate.
Increasing $\kappa$ allows more photons to leak out of the cavity.
On the other hand, decreasing $\kappa$ improves the resolution of the two spectral peaks corresponding to the qubit eigenstates.
Therefore, one anticipates there is an optimal value of the decay rate.
We will determine it in the next subsection.

\subsection{Maximizing SNR with respect to detuning and the pull/damping ratio}
\label{secSNRoptDK}
Here we consider the situation when, apart from the detuning, one is able to vary the ratio of the dispersive pull to the cavity damping rate.
As before, the average number of cavity photons at resonance~\eqref{eqResonantNumberOfPhotons} is maintained constant.
We show how the upper bound on SNR for the measurement can be approached.

The assumption of the variable pull/damping ratio $g\lambda / \kappa$ is quite plausible.
The ratio can be varied in two ways.
First, $\kappa$ can be set on the design stage of an experiment.
Second, for the superconducting experiments, it is usually possible to tune the qubit frequency $\omega_\q$ \emph{in situ}~\cite{clarke2008superconducting} which changes $\lambda$~\eqref{eqLambda}.

Let us introduce the dimensionless detuning and decay rate
\begin{equation}
\label{eqDeltaAndK}
	\Delta = \frac{\omega_{\dr\res}}{g\lambda},
\quad
	K = \frac{\kappa/2}{g\lambda}.
\end{equation}
In these notations the mean counts~\eqref{eqMeanCountExplicit} are given by
\begin{equation}
\label{eqMeanCountDK}
	\overline n_\uparrow = \frac{K^3 \Tm}{(\Delta - 1)^2 + K^2},
\quad
	\overline n_\downarrow = \frac{K^3 \Tm}{(\Delta + 1)^2 + K^2},
\end{equation}
where we have introduced the dimensionless time
\begin{equation}
\label{eqTm}
\Tm = \frac 1 4 \frac{|\alpha^\mathrm{res}|^2}{n_\text{cr}}
	\eta |\omega_\q - \omega_\res| \tm.
\end{equation}
With~(\ref{eqMeanCountDK}-\ref{eqTm}) SNR~\eqref{eqSNRGeneralPoissonian} is expressed as 
\begin{multline}
\label{eqSNRDK}
	\SNR = \sqrt\Tm \, K^{3/2}
\\
	\times\left(
		\frac 1 {\sqrt{(\Delta - 1)^2 + K^2}}
		- \frac 1 {\sqrt{(\Delta + 1)^2 + K^2}}
	\right).
\end{multline}

Equation~\eqref{eqTm} suggests the optical range is favorable for our scheme.
The ratio $|\alpha^\mathrm{res}|^2 / n_\text{cr}$ is small by condition~\eqref{eqDispersive} assuring the measurement is non-demolition.
For the same values of the ratio, bigger difference $|\omega_\q - \omega_\res|$ results in higher $\Tm$ and better readout.
However, by the condition~\eqref{eqRWA} the difference should be much smaller than the characteristic frequencies of the system.
Therefore, high frequencies are favorable.

We find the maximum of SNR with respect to $K$ and $\Delta$.
Equating partial derivatives to zero gives the set of equations
\begin{gather}
\label{eqdDeltaSNR}
	\frac{\Delta + 1}{\big[(\Delta + 1)^2 + K^2 \big]^{3/2}}
	= \frac{\Delta - 1}{\big[(\Delta - 1)^2 + K^2 \big]^{3/2}},
\\
\label{eqdKSNR}
	\frac{3(\Delta + 1)^2 + K^2}{\big[(\Delta + 1)^2 + K^2 \big]^{3/2}}
	= \frac{3(\Delta - 1)^2 + K^2}{\big[(\Delta - 1)^2 + K^2 \big]^{3/2}}.
\end{gather}
Note Eq.~\eqref{eqdDeltaSNR} is the same as~\eqref{eqDOptimal}, despite being written in other notations.
The solution of Eqs.~(\ref{eqdDeltaSNR}-\ref{eqdKSNR}) reads
\begin{equation}
	\Delta = \slashfrac {\sqrt 5} 2,
\quad
	K = \slashfrac {\sqrt 3} 2.
\end{equation}
One can check that, at the $\Delta$ and $K$ given, SNR reaches global maximum.

Using the definitions~\eqref{eqDeltaAndK} and Eq.~\eqref{eqSNRDK}, we express $\omega_{\dr\res}$, $\kappa$, and SNR in terms of $g\lambda$:
\begin{gather}
	\omega_{\dr\res} \approx 1.118 g\lambda,
\quad
	\kappa = 1.732 g\lambda,
\\
\label{eqSNRMaxSymmetrical}
	\SNR \approx 0.570 \sqrt{\eta \tm \, g\lambda |\alpha^\text{res}|^2}.
\end{gather}

Note that using asymmetrical cavity one can increase SNR~\eqref{eqSNRMaxSymmetrical} substantially.
In this case, the overall damping is $\kappa^\bi + \kappa^\bii$, where $\kappa^\bi$ and $\kappa^\bii$ are the rates cavity photons leak through the first and the second port.
If the number of resonator photons is fixed, increasing $\kappa^\bi$ only widens the resonator spectrum;
thus one makes the rate as small as possible.
On the other hand, $\kappa^\bii$ gives the rate photons arrive at the detector.
In the best case of $\kappa^\bi \ll \kappa^\bii$ one changes $\kappa/2$ to $\kappa^\bii$ in the prefactor of~\eqref{eqMeanCount}, while in the other occurrences $\kappa$ is substituted with $\kappa^\bii$.
This doubles the dimensionless time $\Tm$~\eqref{eqTm}.
Meanwhile, the optimal value of $\kappa^\bii$ is the same as the optimal $\kappa$ of the symmetrical case.
We have an increase in SNR by the ratio of $\sqrt 2$.
The resulting value is the upper limit for SNR in the measurement we consider,
\begin{equation}
\label{eqSNRMax}
	\SNR < 0.806 \sqrt{\eta \tm \, g\lambda |\alpha^\text{res}|^2}.
\end{equation}

We have investigated how to achieve the maximum SNR.
However, SNR quantifies the measurement robustness only heuristically.
Note that we haven't even specified the way one distinguishes upper and lower states of the qubit.
Besides, the notion~\eqref{eqSNRGeneral} of SNR takes into account only the two first moments $\overline n$ and $\overline{n^2}$ of the photon count.
SNR can't give a full description of fluctuations that obey Poisson statistics~\eqref{eqP} and thus have non-vanishing moments of higher orders.

In the next section, we consider the measurement in finer detail and give a precise characteristic of its performance.

\section{Fidelity of the thresholding measurement}

In this section, we consider the thresholding measurement of the qubit.
An analytical expression for fidelity of the measurement is given, in terms of the system parameters.
With it, we determine conditions for the maximum fidelity.
Also, it is shown that, for big measurement times, one obtains maximum in fidelity by maximizing SNR.

The easiest way to discriminate the state of the qubit by the number of photocounts is by a \emph{threshold count}, which is set between the counts most probable for each eigenstate.
Then, if the number of detected photons is less than the threshold, the qubit is considered to occupy the ground state;
and it is considered to be in the excited state in the opposite case.

\subsection{Threshold count}
\begin{figure}[ht]
\includegraphics{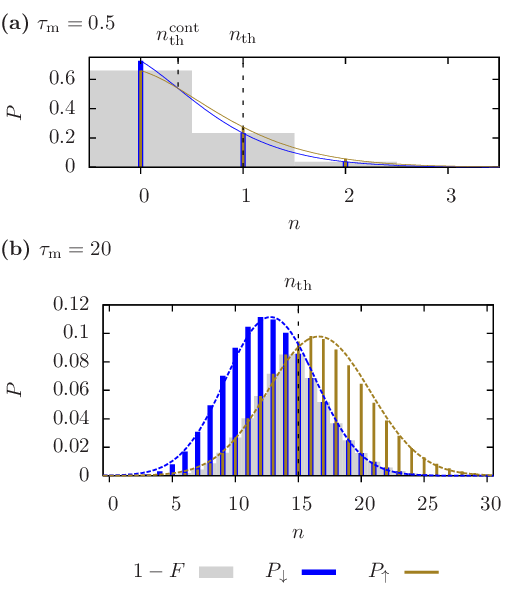}%
\subfigure{%
	\label{figDistributionsHalfaPhoton}%
}%
\subfigure{%
	\label{figDistributions20Photons}%
}%
\caption{(Color online) Probability distributions of photocounts: $P_\uparrow$ for the qubit in the excited state $\ket{\uparrow}$ and $P_\downarrow$ for the qubit in the ground state $\ket{\downarrow}$.
The square of the gray area is equal to the probability of a false measurement result, $1-F$.
For both subfigures, $D = 0.6$ and $X = 0.15$.
Each subfigure is plotted for a measurement time such that:
\subref{figDistributionsHalfaPhoton} Half a photon would be detected on average in case of the drive resonant with the cavity.
Solid lines show the continuation of the Poisson distribution to the real values.
\subref{figDistributions20Photons} Twenty photons would be detected on average in the case of the resonant drive.
Dashed lines show the Gaussian approximation to the distributions.
}
\label{figDistributions}
\end{figure}

It is natural to set the threshold count $n_\thr$ to be the least number of photons detected, for which the probability of the qubit to reside in the upper state $\ket \uparrow$ is bigger than the probability to reside in the lower state $\ket \downarrow$~(see Fig.~\ref{figDistributions}).

To determine $n_\thr$, we first find its continuous analog: the point of intersection $n_\thr^\mathrm{cont}$ of extrapolations of $P_\uparrow(n)$ and $P_\downarrow(n)$ to the real values.
These are given by
\begin{equation}
	P_q^\mathrm{cont} (n)
		= \frac{(\overline n_q)^n \, e^{-\overline n_q}}{\Gamma(n+1)},
\quad
	q = \, \uparrow, \downarrow.
\end{equation}
In other words, $n_\thr^\mathrm{cont}$ is defined by
\begin{equation}
	P_\uparrow^\mathrm{cont}(n_\thr^\mathrm{cont})
		= P_\downarrow^\mathrm{cont}(n_\thr^\mathrm{cont}).
\end{equation}
The solution of the equation is
\begin{equation}
\label{eqnthrcont}
	n_\thr^\mathrm{cont} = \frac{\overline n_\uparrow - \overline n_\downarrow}
				{\log\overline n_\uparrow - \log\overline n_\downarrow}.
\end{equation}
Plots of continuous distributions $P_{\uparrow,\downarrow}^\mathrm{cont}$ are given in Fig.~\ref{figDistributionsHalfaPhoton}, with the positions of $n_\thr^\mathrm{cont}$ and $n_\thr$ marked.

The threshold count $n_\thr$ can be found from $n_\thr^\mathrm{cont}$.
The peak of $P_\uparrow^\mathrm{cont}$ is higher and is located to the right of that of $P_\downarrow^\mathrm{cont}$.
Also, as~\eqref{eqnthrcont} shows, there is only one point where the probabilities are equal.
It follows then, from graphical considerations, that $P_\uparrow^\mathrm{cont}(n) > P_\downarrow^\mathrm{cont}(n)$ for $n > n_\thr^\mathrm{cont}$.
Therefore,
\begin{equation}
\label{eqnthr}
	n_\thr = \lceil n_\thr^\mathrm{cont} \rceil
		= \left\lceil 
			\frac{\overline n_\uparrow - \overline n_\downarrow}
					{\log\overline n_\uparrow - \log\overline n_\downarrow}
		\right\rceil,
\end{equation}
where $\lceil x \rceil$ denotes the first integer not less than $x$.

\subsection{Fidelity}

Fidelity is related to the probability of correct measurement result.
If the number of detected photons is not less than the threshold count $n_\thr$~\eqref{eqnthr} and one determines the qubit to be in the upper state, there is still a probability that the qubit is in the lower state, and vice versa [see Fig.~\ref{figDistributions20Photons}].
Fidelity is expressed as
\begin{equation}
\label{eqFidelityDefinition}
	F = 1 - \sum_{n=0}^{n_\thr - 1} P_\uparrow(n)
		- \sum_{n=n_\thr}^\infty P_\downarrow(n).
\end{equation}

The sums in the definition can be expressed in terms of the incomplete Gamma function, 
\begin{equation}
\label{eqIncompleteGamma}
\Gamma(n,x) = \int_x^\infty dt\, t^{n-1} e^{-t}.
\end{equation}
It is shown in Appendix~\ref{apIncompleteGamma}, that
\begin{equation}
\label{eqGamma}
	\Gamma(n,x) = e^{-x} (n-1)! \sum_{n=0}^{n-1} \frac{x^n}{n!}
\end{equation}
for $n$ integer.
We re-express the second sum in~\eqref{eqFidelityDefinition},
\begin{equation}
	\sum_{n=n_\thr}^\infty P_\downarrow(n)
		= 1 - \sum_{n=0}^{n_\thr - 1} P_\downarrow(n),
\end{equation}
and then, apply~\eqref{eqGamma} to each sum with the probabilities given by~\eqref{eqP}.
One arrives at
\begin{equation}
\label{eqFidelity}
	F = \frac{\Gamma(n_\thr, \overline n_\downarrow)
			- \Gamma(n_\thr, \overline n_\uparrow)}
		{\Gamma(n_\thr)}.
\end{equation}

As a special case, the fidelity of a click/no-click detector can be obtained from~\eqref{eqFidelityDefinition}.
Such a detector clicks (with a probability $\eta$) when at least one photon is absorbed.
In this case, we decide that the qubit is in the upper state.
In the opposite case, if no photons were detected during the measurement, one decides that the qubit is in the lower state.
The situation is captured by $n_\thr = 1$.
Then~\eqref{eqFidelityDefinition} reduces to
\begin{equation}
\label{eqFidelityOnOff}
	F_\onoff = e^{-\overline n_\downarrow} - e^{-\overline n_\uparrow}.
\end{equation}

In Ref.~\cite{govia2014high}, fidelity for the related measurement with the on/off detector was obtained;
the quantity is called the measurement contrast there.
Formally, expression~\eqref{eqFidelityOnOff} coincides with that of the mentioned work, when the latter is taken in the limit of negligible number of dark counts.
However, a different measurement sequence is considered in the reference.
There, a photon is allowed to leave the resonator only after the cavity has been already pumped.
This is in contrast to the case of continuous drive and permanent leakage considered in the present paper.

One can consider how the choice of detuning maximizing SNR improves fidelity, as compared to the naive choice $D = X$.
This can be done using the expression for fidelity~\eqref{eqFidelity} and the numerical solutions of Eq.~\eqref{eqDOptimal}.
First, we find there is no need to use a detuning other than $D = X$ for stronger measurements beginning with $X=1$.
This was already shown by analyzing SNR in Sec.~\ref{secSNRoptDX}. 
In the case of smaller $X$, consider a measurement reaching 95\% fidelity.
For $X=0.5$, one would need the measurement to last 1.5 times less if the detuning maximizing SNR is used.
And for $X=0.15$ such choice of detuning shortens the measurement more than sevenfold!
Given such performance, it is natural to pose certain questions:
Is there some connection between the conditions of maximum of SNR and fidelity?
Would there be any further advantage in using the detuning \emph{maximizing fidelity}?
We address these questions below.

\subsection{Gaussian approximation}
\label{secGaussian}

Consider a Gaussian approximation to the threshold count and fidelity.
We will show that in this approximation fidelity is expressed in terms of SNR.

One obtains the approximation as follows.
For a long measurement and lots of photocounts,
\begin{equation}
\label{eqGaussialityCondition}
	\overline n_\uparrow, \overline n_\downarrow \gg 1,
\end{equation}
Poisson distributions $P_\uparrow$ and $P_\downarrow$ are well-approximated with Gaussians:
\begin{equation}
\label{eqGaussianP}
	P_q(n) \approx \frac 1 {\sqrt{2\pi \overline{\Delta n^2}_q}}
					\exp\left[-\frac{(n - \overline n_q)^2}
									{2 \overline{\Delta n^2}_q}
					\right],
\quad
	q = \, \uparrow, \downarrow.
\end{equation}
The approximation is shown in Fig.~\ref{figDistributions20Photons}.
With it, fidelity~\eqref{eqFidelityDefinition} is expressed as
\begin{equation}
\label{eqGaussianFidelity}
	F \approx \frac 1 {\sqrt\pi} \int_{x_\uparrow}^{x_\downarrow} dx \, e^{-x^2}
\\
	= \frac 1 2 \erf x_\downarrow - \frac 1 2 \erf x_\uparrow,
\end{equation}
where the limits of integration are given by
\begin{equation}
\label{eqGaussianLimits}
	x_q = \frac{n_\thr^\mathrm{gauss} - n_q}{\sqrt{2 \overline{\Delta n^2}_q}},
\quad
	q = \, \uparrow, \downarrow,
\end{equation}
and the Gaussian threshold count is
\begin{equation}
\label{eqGaussianThresold}
	n_\thr^\mathrm{gauss} = \sqrt{\overline n_\uparrow \overline n_\downarrow
		\left(
			1 + \frac{\log \overline n_\uparrow - \log \overline n_\downarrow}
					{\overline n_\uparrow - \overline n_\downarrow}
		\right)}
	\approx \sqrt{\overline n_\uparrow \overline n_\downarrow}.
\end{equation}
To derive the approximate identity in~\eqref{eqGaussianThresold}, it was taken into account that $n_\thr$~\eqref{eqnthr} is huge compared to unity due to the condition~\eqref{eqGaussialityCondition}.
Also, we have used the error function
\begin{equation}
	\erf x = \frac 2 {\sqrt\pi} \int_0^x dx' e^{-x'^2}
\end{equation}
to express the integral in~\eqref{eqGaussianFidelity} in a convenient form.

Using~\eqref{eqGaussianThresold}, one expresses the parameters of the error function in~\eqref{eqGaussianFidelity} in terms of SNR~\eqref{eqSNR}:
\begin{gather}
\label{eqErrorFnParameters}
	x_\uparrow \approx -\SNR/\sqrt 2,
\quad
	x_\downarrow \approx \SNR/\sqrt 2.
\end{gather}
The detuning which maximizes Gaussian fidelity coincides with that maximizing SNR.

Equations~(\ref{eqGaussianFidelity},~\ref{eqErrorFnParameters}) constitute the expression for fidelity which is formally equivalent to that given in Ref.~\cite{gambetta2007protocols}.
To show this, one should change our definition of SNR in accordance to the reference.
There, SNR is defined as a ratio of signal and noise \emph{powers}.
Thus, to arrive at the formula given in Ref.~\cite{gambetta2007protocols}, one should square the r.h.s. of the SNR definition~\eqref{eqSNRGeneral}.
Still, we have derived the expression for a different case:
The distributions~\eqref{eqGaussianP} are of Poissonian width set by Eq.~\eqref{eqMeanSquaredCount} contrary to the same-width case considered in the reference;
The distributions are sufficiently narrow, by Eqs.~\eqref{eqGaussialityCondition} and \eqref{eqMeanSquaredCount}.

We have shown that the maxima of SNR and fidelity coincide for long measurement times.
Next we are going to investigate the exact conditions for maximum fidelity.

\subsection{Maximizing fidelity with respect to detuning}

In this subsection, we find the detuning that maximizes measurement fidelity.
Here the duration of measurement, the dispersive pull, and the resonator decay  rate are fixed. 
Just as before, we assume the average number of cavity photons at resonance~\eqref{eqResonantNumberOfPhotons} is maintained constant.

\begin{figure}[ht]
\includegraphics{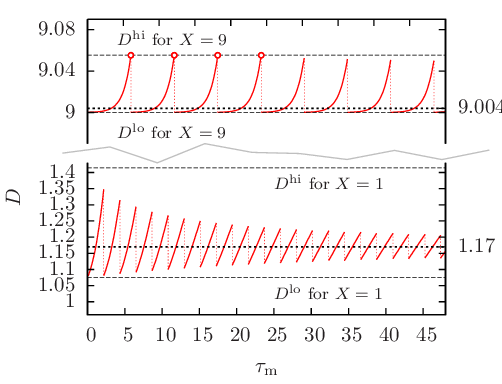}
\caption{(Color online) Dependence of the dimensionless optimal detuning on the dimensionless measurement duration, for a fixed dispersive pull and resonator leakage:
	the detuning maximizing fidelity for $X=9$~(top red solid) and $X=1$~(bottom red solid),
	the detuning maximizing SNR~(bold dotted),
	the upper and lower bounds on the detuning~(dashed).
The lower bound is not reached, despite the way it looks.
The points marked with empty circles are not reached too.
}
\label{figDmax}
\end{figure}

The threshold count~\eqref{eqnthr} changes in steps with the measurement duration.
One then expects that the optimal detuning has discontinuities in points where the threshold changes.
This behavior is illustrated in Fig.~\ref{figDmax}.

Let us consider durations $\taum$ between the discontinuities.
Here, $n_\thr$ is constant; points where fidelity $F$ has extrema are found by equating the first derivative of $F$ to zero, $\at{\slashfrac{\partial F}{\partial D}}{n_\thr \equiv \const} = 0$.
Carrying out the derivative of~\eqref{eqFidelity}, the condition for extremum resolves to
\begin{equation}
\label{eqMaxFCondition}
	\frac{\Dopt+X}{\Dopt-X} 
		\, e^{\overline n_\uparrow-\overline n_\downarrow}
	= \left(\frac{\overline n_\uparrow}
				{\overline n_\downarrow} \right)^{n_\thr+1}.
\end{equation}
It is shown in Appendix~\ref{apMaxF} that the extrema given by the equation are the points of maximal fidelity.
The solution of Eq.~\eqref{eqMaxFCondition} gives the optimal detuning between two `jumps'.

Abrupt change of $\Dopt$ takes place in two cases.
First, it occurs when the fidelity for the next threshold count $n_{\thr}+1$ exceeds that for the current one, $n_\thr$.
Strictly speaking, as $F(\taum)$ is continuous, at the point of `jump' there exist two values of $\Dopt$.
This is the case for the curve with $X = 1$ in Fig.~\ref{figDmax}, and the transitions after the fourth one for the curve with $X=9$.
On the other hand, beginning with some measurement duration $\taum$, there can be no detuning satisfying~\eqref{eqMaxFCondition} at the present $n_\thr$.
This case is realized in the first four transitions occurring for $X=9$, as illustrated in Fig.~\ref{figDmax}.
In both cases, $n_\thr$ increments by one and $\Dopt$ switches to the value given by~\eqref{eqMaxFCondition} with the threshold incremented.

The last case allows us to find upper and lower bounds on oscillations of the optimal detuning.
Consider the interval between two `jumps'.
On the interval $\Dopt$ changes continuously, governed by~\eqref{eqMaxFCondition}; with it, the continuous threshold $n_\thr^\mathrm{cont}$~\eqref{eqnthrcont} changes.
According to the definition~\eqref{eqnthr} of the threshold count, $n_\thr$ increases when $n_\thr^\mathrm{cont}$ passes by the current threshold.
The value $\Dopt$ reaches prior to the increment is the highest possible.
One can find this highest possible value.
We substitute $n_\thr$ in~\eqref{eqMaxFCondition} with $n_\thr^\mathrm{cont}$, given by~\eqref{eqnthrcont}.
The substitution gives rise to the equation
\begin{equation}
	(D+X)\big[(D-X)^2 + 1\big] = (D-X)\big[(D+X)^2 + 1\big].
\end{equation}
Solution of the equation sets the upper bound on $\Dopt$:
\begin{equation}
\label{eqDhi}
	\Dopt \le \Dhi,
\quad
	\Dhi = \sqrt{X^2 + 1}.
\end{equation}
As at the point of `jump' $n_\thr = n_\thr^\mathrm{cont} + 1$, the value $\Dopt$ approaches to after the switch can be found in an analogous way.
We substitute $n_\thr$ in~\eqref{eqMaxFCondition} with $n_\thr^\mathrm{cont} + 1$, which results in the equation
\begin{equation}
	(D+X)\big[(D-X)^2 + 1\big]^2 = (D-X)\big[(D+X)^2 + 1\big]^2.
\end{equation}
The equation reduces to the quartic equation.
It has one real positive root.
The root is the lower bound on $\Dopt$:
\begin{equation}
\label{eqDlo}
	\Dopt > \Dlo,
\quad
	\Dlo = \sqrt{\nicefrac 1 3 \, (2\sqrt{X^4 + X^2 + 1} + X^2 - 1)}.
\end{equation}
Knowledge of the upper and lower bounds speeds up dramatically the numerical procedures to obtain $\Dopt$.

We briefly review the numerical procedures used.
The simplest way to determine $\Dopt$ is to calculate $F$ for $D$ changing from $\Dlo$ to $\Dhi$ with small steps, and choose the detuning resulting in the biggest fidelity.
The plot for $X=1$ in Fig.~\ref{figDmax} was obtained this way.
In the calculations, the interval from $\Dlo$ to $\Dhi$ was divided in 1000 steps.
However, for $X=9$, variations of $F$ are too small to use this method.
The respective curve in Fig.~\ref{figDmax} was calculated by solving~\eqref{eqMaxFCondition} for each $\taum$.
In both cases, $\taum$ was changing with a step of $0.1$.
As $\Dopt$ is quite steep just before a `jump', we determine each point of a `jump' as precise as possible.
In the case $\Dopt$ reaches $\Dhi$, a switch occurs when $n_\thr^\mathrm{cont}$ becomes equal to the current threshold.
In the other case, the time is found using the fact that fidelities for $n_\thr$ and $n_\thr + 1$ are equal at the time of a switch.

\subsection{Maximizing fidelity with respect to detuning and the pull/damping ratio: the upper bound on the measurement fidelity}
\label{secDKFidelity}

Here we give the detuning and the ratio of dispersive pull to resonator leakage which result in optimal fidelity.
The average number of cavity photons at resonance is kept fixed.
The resulting fidelity is the biggest possible fidelity for the scheme considered.
Using the optimal parameters, we show the possibility for high-fidelity single-shot readout in various realizations of the scheme.

\begin{figure}[ht]
\includegraphics{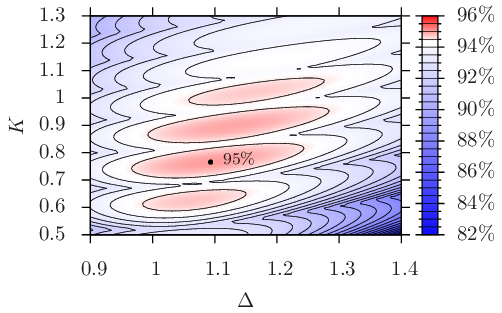}
\caption{(Color online) Fidelity at dimensionless time $\Tm = 11.29$ vs. the dimensionless detuning $\Delta$ and the dimensionless resonator leakage $K$.}
\label{figFidelityDK}
\end{figure}

We use the expressions of mean counts~\eqref{eqMeanCountDK} in terms of $\Delta$ and $K$~\eqref{eqDeltaAndK}.
A typical dependence of fidelity~\eqref{eqFidelity} on those parameters is given in Fig.~\ref{figFidelityDK}.

One can write out the equations on stationary point of fidelity.
Carrying out the partial derivatives $\slashfrac{\partial F}{\partial\Delta}$ and $\slashfrac{\partial F}{\partial K}$ and equating them to zero gives rise to the set of equations:
\begin{gather}
\label{eqDeltaExtremal}
	\frac{\Delta + 1}{\Delta - 1}
		\, e^{\overline n_\uparrow-\overline n_\downarrow}
	= \left(\frac{\overline n_\uparrow}
				{\overline n_\downarrow} \right)^{n_\thr+1},
\\
\frac{3(\Delta + 1)^2 + K^2}{3(\Delta - 1)^2 + K^2}
		\, e^{\overline n_\uparrow-\overline n_\downarrow}
	= \left(\frac{\overline n_\uparrow}
				{\overline n_\downarrow} \right)^{n_\thr+1}.
\end{gather}
As before, $n_\thr$ is assumed constant during the differentiation.
It follows from the equations that 
\begin{equation}
\label{eqKinTermsOfDelta}
	K^2 = 3\Delta^2 - 3,
\quad
	\Delta \ne \pm 1.
\end{equation}
With this, the mean counts~\eqref{eqMeanCountDK} can be expressed in terms of $\Delta$ alone:
\begin{gather}
\label{eqnUpInTermsOfDelta}
	\overline n_\uparrow = \Tm \frac{\sqrt{27}}2
				\frac{(\Delta + 1)\sqrt{\Delta^2 - 1}}{2\Delta + 1},
\\
\label{eqnDownInTermsOfDelta}
	\overline n_\downarrow = \Tm \frac{\sqrt{27}}2
				\frac{(\Delta - 1)\sqrt{\Delta^2 - 1}}{2\Delta - 1}.
\end{gather}

\begin{figure}[ht]
\includegraphics{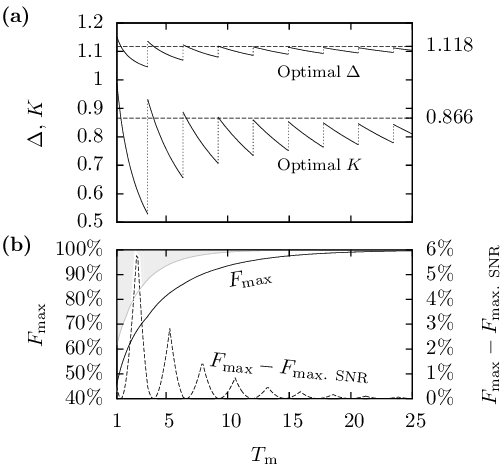}%
\subfigure{%
	\label{figOptDK}%
}%
\subfigure{%
	\label{figDKFdiff}%
}%
\caption{\subref{figOptDK} Top curves: the dimensionless detuning $\Delta$ maximizing fidelity (solid curve) and  SNR~(dashed line).
	Lower curves: dimensionless decay rate $K$ of the resonator that maximize fidelity (solid curve) and SNR~(dashed line).
	\subref{figDKFdiff} Maximal fidelity for symmetrical resonator~(black solid), maximal fidelity for asymmetrical resonator~(gray solid), and the difference between the optimal fidelity and the fidelity reached with the parameters maximizing SNR~(dashed).
}
\label{figOptDKandF}
\end{figure}

Plots of optimal $\Delta$ and $K$ in Fig.~\ref{figOptDK} are obtained with Eqs.~(\ref{eqDeltaExtremal}, \ref{eqKinTermsOfDelta}-\ref{eqnDownInTermsOfDelta}).
This is done analogously to the calculation of $X=9$ plot in Fig.~\ref{figDmax}.
The key difference is that Eqs.~\eqref{eqnUpInTermsOfDelta} and \eqref{eqnDownInTermsOfDelta} are used to calculate fidelity~\eqref{eqFidelity}.
Also, here we have numerically checked that the computed critical points are the points of maxima.

The asymmetrical resonator with a negligible first-port leakage can be tackled in the same way as in Sec.~\ref{secSNRoptDK}.
It was shown that the dimensionless time $\Tm$~\eqref{eqTm} is multiplied by the factor of two in this case, so that a given fidelity is reached quicker.
This gives the upper bound on fidelity for the measurement with various ratios of port leakage rates.
The bound is shown in Fig.~\ref{figDKFdiff}, as well as the fidelity for the case of symmetrical resonator.
Numerical estimations for the measurement duration are given in Table~\ref{tabEstimations}.

\begin{table}
\caption{\label{tabEstimations}Estimations for measurement time for different realizations of cavity QED: transmon superconducting qubit coupled to 1D resonator, quantum dot~(QD) in a nanocavity, Bose-Einstein condensate~(BEC) in a 3D optical cavity, and an ensemble of nitrogen defect spins in diamond coupled to a 1D microwave resonator.
Efficiency of the detector is $\eta = 0.9$.}
\newcommand\T{\rule{0pt}{2.6ex}}       
\newcommand\B{\rule[-1.2ex]{0pt}{0pt}} 

\newcommand\he[1]{\multicolumn{1}c{#1}} 
\newcommand\hetwo[1]{\multicolumn{2}c{#1}}
\newcommand\hel[1]{\multicolumn{1}l{#1}} 

\newcolumntype{.}{D{.}{.}{3}}
\newcolumntype{t}{D{.}{.}{1}}

\begin{ruledtabular}
\begin{tabular}{lccc..r@{}@{}l}
	\hel{Qubit} & \he{$g$\footnotemark[1]}
		& \he{$|\omega_\q\!-\!\omega_\res|$}
		& \he{$|\alpha^\mathrm{res}|^2$}
		& \hetwo{$\tm$ for $F$ of}
		& \hetwo{$T_1$\footnotemark[1]} \\
	& \he{(\unit{MHz})} 
		& \he{(\unit{GHz})}
		&& \he{95\%} & \he{99\%}\B\\
	\hline
	\T Transmon \cite{jeffrey2014fast}
		& 86 & 1\footnotemark[1] & 1\footnotemark[1] 
		& \unit[0.7]{\mu s} & \unit[1.2]{\mu s}
		& 20 & \unit{$\mu$s}\\
	QD \cite{hennessy2007quantum}
		& 21000 & 1000 & 10
		& \unit[1.2]{ns} & \unit[2.0]{ns}
		& 11 & \unit{ns}\\
	BEC \cite{brennecke2007cavity}
		& 1000 & 1000 & 1000 
		& \unit[5.1]{ns} & \unit[11.0]{ns}
		& 53 & \unit{ns}\\
	N de-\\ fects \cite{ranjan2013probing}
		& 17 & 0.1 & 0.01
		& \unit[17.6]{\mu s} & \unit[30.6]{\mu s}
		& 20 & \unit{s}\\
\end{tabular}
\end{ruledtabular}
\footnotetext[1]{Value from reference.}

\end{table}

It is important to use the photon-number-resolving detector to achieve a high-fidelity readout.
The on/off detector provides the optimal fidelity only up to the first `jump' in parameter values.
After the `jump', the optimal fidelity is achieved with $n_\thr \geq 2$, which is impossible for the on/off detector.
The possibility to resolve a photon number matters starting from a fidelity of about 73\%.

Note the naive detuning ($\Delta = 1$) and the optimal $K$ result in a non-substantial loss of fidelity.
The loss is below 1\% for a fidelity higher than 75\%.
That is not surprising.
We have already discussed in Sec.~\ref{secSNRoptDX} that the naive detuning provides a close-to-optimal fidelity for $g\lambda > \kappa/2$~($K < 1$).
One can see from Fig.~\ref{figOptDKandF} that the optimal $K$ satisfies this condition for any reasonable value of the fidelity.

\section{Use of the parameters maximizing fidelity and those optimizing SNR}

In this section, we discuss how using the parameters maximizing fidelity compares to the use of those maximizing SNR, in terms of the resulting fidelity.

First, for big measurement times, maximizing SNR results in the maximum of fidelity.
We have shown this in Sec.~\ref{secGaussian}, using the Gaussian asymptotics to fidelity.

For smaller times, it depends on whether one is free to choose only the drive-resonator detuning, or both the detuning and the resonator decay rate.

Consider the case one is able to choose a detuning only.
For this case, we have performed a numerical comparison of the fidelity $F_\text{max}$ reached with the optimal detuning, and the fidelity $F_\text{max. SNR}$ the detuning maximizing SNR results in.
A non-vanishing difference $F_\text{max} - F_\text{max. SNR}$ occurs near the point of threshold change.
It is found that the maximal difference is slightly bigger than 1\%.
Such a gain is achieved for pulls $X \approx 0.5 \div 1$.
The fidelities reached under these circumstances are about $50\% \div 60\%$.

If, in addition to detuning, it is possible to set the dispersive pull or the resonator leakage, the gain becomes bigger.
As shown in Fig.~\ref{figDKFdiff}, use of the parameters maximizing fidelity rather than SNR pays off with a moderate increase of fidelity, for a range of measurement durations.
Namely, a gain about 6\% occurs in the measurement reaching fidelity about 65\%.
And even for fidelity of about 95\% the increase can be close to 1\%.

\section{Conclusion}

We have determined the optimal system parameters for the dispersive qubit readout using a fully-absorbing, photon-number-resolving detector.

The drive-resonator detuning equal to the dispersive pull may result in suboptimal measurement performance.
Namely, this is the case for a weak measurement, when the dispersive pull is smaller than the cavity decay rate, $g\lambda < \kappa/2$.
Consider the case only the detuning can be varied.
Then, the optimal detuning can be determined by Fig.~\ref{figMaxSNR}.
For a very weak measurement, $g\lambda \ll \kappa/2$, we have obtained asymptotics for the optimal detuning: $\omega_{\dr\res} = \kappa \sqrt2/4$.

To obtain these results, it is sufficient to maximize a simple characteristic of the measurement, SNR, which is given by~\eqref{eqSNRGeneral}.
It turns out that the detuning maximizing fidelity~\eqref{eqFidelityDefinition} results in almost the same values of fidelity.
For sufficiently long measurement durations that result in high fidelity, we have proved that the conditions of maxima of SNR and fidelity coincide.
As for the moderate fidelities, the difference in fidelity is not substantial.

The situation is different if one is able to tune the $g\lambda/\kappa$ ratio.
One can use the ratio that maximizes SNR if aiming at more than $95\%$ fidelity.
A fidelity very close to the optimal one is then achieved, while the ratio is the same for measurements of any duration.
For a shorter measurement, it is better to use $g\lambda/\kappa$ that maximizes fidelity.
It is given in Fig.~\ref{figOptDK}.
For each measurement duration, there is a distinct $g\lambda/\kappa$ ratio.
(This does not mean the ratio should be changed throughout the measurement.)
As for the drive-resonator detuning, it can be chosen as $\omega_{\dr\res} = g\lambda$.
This results in almost the same fidelity as the exact value of optimal detuning.

The photon-number-resolving detector is advantageous for the readout.
With a click/no-click detector, one needs a longer measurement to achieve fidelities starting with 72\%.

Single-shot readout using the considered scheme is achievable in various cavity-QED type systems~(see Table~\ref{tabEstimations}).
The readout can reach 99\% fidelity.
This opens the possibility of using the readout in quantum error-correction schemes.
Note our scheme is best-suited for high frequencies of both the qubit and the cavity, as follows from~\eqref{eqTm} and our comments on it.

Our results apply not only to a single-shot measurement but to the sequential measurements as well.
In this case the measurement time discussed is replaced with the sum of durations of all measurements in a sequence.

\begin{acknowledgments}
The author thanks A.~Semenov and O.~Chumak for useful discussions, and E.~Stolyarov for critical reading of the manuscript.
\end{acknowledgments}

\appendix

\section{Cumulative distribution function of Poisson process}
\label{apIncompleteGamma}

Here, we express the cumulative distribution function of Poissonian process in terms of incomplete Gamma function.

The probability that a Poisson random variable $\xi$ occurs with a value less than or equal to $N$ is
\begin{equation}
\label{eqPoissonCDF}
	P(\xi \le N) = \sum_{n=0}^N \frac{\lambda^n}{n!} e^{-\lambda}.
\end{equation}
This quantity is known as cumulative distribution function of the variable.
Incomplete Gamma function is defined with the following expression:
\begin{equation}
\label{eqGammaUpper}
	\Gamma(n,x) = \int_x^\infty dt\, t^{n-1} e^{-t}.
\end{equation}
A basic property of the incomplete Gamma function reads
\begin{equation}
\label{eqGammaUpperRecurrecnceRelation}
	\Gamma(n+1,x) = n\Gamma(n,x) + x^n e^{-x}.
\end{equation}
It can be derived integrating by parts the definition~\eqref{eqGammaUpper}.
Noticing that
\begin{equation}
	\Gamma(1,x) = e^{-x}
\end{equation}
and applying induction to~\eqref{eqGammaUpperRecurrecnceRelation}, one arrives at
\begin{multline}
\label{eqGammaUpperInSeries}
	\Gamma(n+1,x) = e^{-x} \sum_{k=0}^n x^k n(n-1) \ldots (n-k+1)
\\
		= e^{-x} n! \sum_{k=0}^n \frac{x^k}{k!}.
\end{multline}
In obtaining~\eqref{eqGammaUpperInSeries} it was assumed that $n$ is a positive integer or zero.
With~\eqref{eqGammaUpperInSeries}, the cumulative distribution function~\eqref{eqPoissonCDF} is expressed as follows: 
\begin{equation}
	P(\xi \le N) = \frac {\Gamma(N+1, \lambda)} {\Gamma(N+1)},
\quad
	N = 0, 1, \ldots
\end{equation}

\section{Conditions for fidelity maximum at the extremal point}
\label{apMaxF}

In this appendix, we show that the extrema $\Dopt$ given by~\eqref{eqMaxFCondition} are the maxima of fidelity.

First of all, the highest $\Dopt$ possible, $\Dhi$~\eqref{eqDhi}, maximizes fidelity $F$.
This follows from the fact that the derivative $\at{\slashfrac{\partial F}{\partial D}}{n_\thr^\mathrm{cont} \equiv \const}$ changes its sign from plus to minus while $\Dopt$ passes $\Dhi$.

We now check the solutions of~\eqref{eqMaxFCondition} give maxima in the rest of the region between two `jumps', i.e., for $\Dopt < \Dhi$.
In this region, extremum of $F$ is a maximum if
\begin{equation}
\label{eqReallyMaximum}
	\at{\frac{\partial^2 F}{\partial D^2}}{n_\thr \equiv \const} < 0.
\end{equation}
Performing differentiation, one obtains
\begin{multline}
\label{eqBigUglyDemonInTheMiddleOfTheRoad}
	\frac{(D+X)^2 + 1 + 2(\taum - n_\thr - 1)(D+X)}
		{\big[(D+X)^2 + 1\big]^{n_\thr + 2}}
	\cdot e^{-\overline n_\downarrow}
\\
	< \frac{(D-X)^2 + 1 + 2(\taum - n_\thr - 1)(D-X)}
		{\big[(D-X)^2 + 1\big]^{n_\thr + 2}}
	\cdot e^{-\overline n_\uparrow}.
\end{multline}
For $X>0$ the stronger inequality can be obtained by multiplying the denominator of the right-hand side of~\eqref{eqBigUglyDemonInTheMiddleOfTheRoad} by $[(D-X)^2 + 1]/[(D+X)^2 + 1]$.
Simplifying the resulting inequality using~\eqref{eqMaxFCondition} and taking logarithm of both sides of it, we have
\begin{equation}
\label{eqBoundOnDiff}
	\overline n_\uparrow - \overline n_\downarrow 
		< n_\thr \log\frac{\overline n_\uparrow}{\overline n_\downarrow}.
\end{equation}
Using the definition~\eqref{eqnthrcont} of $n_\thr^\mathrm{cont}$,~\eqref{eqBoundOnDiff} reduces to
\begin{equation}
	n_\thr^\mathrm{cont} < n_\thr
\end{equation}
which is identity for $\Dopt < \Dhi$ due to the definition of $n_\thr$~\eqref{eqnthr}.

\bibliography{common_sources,jj_sources,cqed,measurement,counters,array,comp}
\bibliographystyle{apsrev4-1}
  
\end{document}